\documentclass[twocolumn,letterpaper,amsmath,amssymb,amsfonts,floatfix,aps,superscriptaddress,showkeys,showpacs,nofootinbib]{revtex4-1}
 
\usepackage{hyperref} 
\usepackage{verbatim} 
\usepackage{graphicx}
\usepackage{bm} 
\usepackage{mathtools}
\usepackage{ifpdf} 
\usepackage{dcolumn}  
\usepackage{epsfig} 
\usepackage{color}
\usepackage[usenames,dvipsnames]{xcolor} 
\usepackage{multirow}
\usepackage{float}
\usepackage[makeroom]{cancel} 

\def\bnabla{\mbox{\boldmath $\nabla $}}

\begin{document}

\title[General mean-field formalism for electrostatic double layers with non-electrostatic bulk and surface interactions]{General theory of charge regulation and surface capacitance}

\author{Rudolf Podgornik}
\email{podgornikrudolf@ucas.ac.cn}
\affiliation{School of Physical Sciences and Kavli Institute for Theoretical Sciences, University of Chinese Academy of Sciences, Beijing 100049, China and CAS Key Laboratory of Soft Matter Physics, Institute of Physics, Chinese Academy of Sciences, Beijing 100190, China}
\affiliation{Department of Theoretical Physics, Jo\v zef Stefan Institute, SI-1000 Ljubljana, Slovenia and Department of Physics, Faculty of Mathematics and Physics, University of Ljubljana, SI-1000 Ljubljana, Slovenia}

\begin{abstract}
A generalization of the mean-field approach will be derived that will take into account the ion-ion as well as ion-surface non-electrostatic effects on an equal footing, being based on the bulk and surface equations of state in the absence of electrostatic interactions. This approach will be applied to the analysis of a single planar surface with dissociable sites with several models of the specific ion-surface non-electrostatic interactions, providing a general thermodynamic insight into the characteristics of the surface differential capacitance. The packing considerations due to surface adsorbed ions will be shown to be more relevant then the bulk packing constraints for ions vicinal to the surface, and to set in prior to the conditions where the bulk packing constraints would become relevant.
\end{abstract}

\maketitle

\section{Introduction}

Short-range, non-electrostatic interactions between ions in solution and/or between ions and the bounding interfaces in aqueous electrolytes as well as ionic liquids under confinement are receiving renewed interest \cite{Perkin2012, Kornyshev-Fedorov} as detailed experiments are yielding results that can not be aligned with the Gouy-Chapman double layer framework and theories are venturing beyond the electrostatic double layer paradigm \cite{Kornyshev-rev,Ryan2010,Yochelis1,Gavish1,Seddon2018}. These short-range interactions are standardly associated with van der Waals,  steric, and hydration effects, and could be in addition coupled to the local dielectric inhomogeneities as well as possibly non-local dielectric effects \cite{Woodward2010,Kornyshev-Fedorov,Forsman2018, Uematsu}.  While detailed atomistic \cite{Laaksonen2018} and coarse-grained Monte Carlo (MC) \cite{Forsman,LevinDC} and Molecular Dynamics (MD) simulations \cite{Holm2009, Holm} are probably our best bet to understand the quantitative details of the behavior of confined general Coulomb fluids, simple analytical theories \cite{Kornyshev-Fedorov, Bazant2, Yochelis1,Gavish1, Kornyshev2018} and maybe even more the proper thermodynamic conceptual framework \cite{maggs2016general, Blossey}, would be helpful - as it was traditionally in the past - to cathegorize the complicated observed phenomena within some simple phenomenological frameworks.

Recently it was shown that simple, local thermodynamic description of any inhomogeneous fluid with a known bulk equation of state, can be generalized to the case of the same system, but charged up with added electrostatic interactions between all mobile particles, on top of all the other interactions that are determining its uncharged equation of state \cite{maggs2016general}. The main point of this generalized mean-field approach is that one does not need to know all the details of the non-electrostatic part of the inter-particle interactions, and the ensuing bulk homogeneous thermodynamic equation of state actually completely suffices, possibly augmented either by non-linear density couplings \cite{Leermalers2009} or by the second order density gradient expansion \cite{Blossey}. This approach is first of all applicable to the {\sl trivial case} of background ideal gas, yielding expectedly the Poisson-Boltzmann (PB) description, but also providing generalizations of the PB theory in the case of different lattice gas models and then all the way to the Carnahan-Stirling fluid and possibly other even more sophisticated models of the non-electrostatic background, yielding in each case a {\sl generalized Poisson-Boltzmann theory}. This framework, based on straightforward thermodynamic identities provided by the Legendre transform of the free energy function of the uncharged system with imposed Gauss theorem for the electrostatic part, opens up also other possible advances to characterize the non-electrostatic effects at the bounding surfaces of the charged system. 

As will become clear, bounding surfaces in general contribute a Gibbsian surface part to the bulk part of the total thermodynamic potential, representing specific non-electrostatic as well as electrostatic interactions at the interface \cite{Hill, Widom}. Since the surface part of the thermodynamic potential after minimization with the bulk part yields a boundary condition for the charge at the interface \cite{Schwinger}, the inclusion of the surface part to the full thermodynamic description in general describes a {\sl charge regulation mechanism} operating at the interface, connecting the surface charge with the surface potential \cite{ninham, Prieve1976, chan1976electrical, chan1983, carnie1993, borkovec1999general, borkovec2001ionization, Netz2003,trefalt2016charge}. The total thermodynamic potential composed of the bulk equation of state and the surface equation of state containing all the non-electrostatic parts of the ion-ion and ion-surface interactions, augmented by the Gauss law for the bulk and for the surface, then yield a complete mean-field description for any system whose non-electrostatic bulk and surface equations of state are known or can be modeled. While this is certainly not the ultimate solution of the problem it does create a consistent and common background to analyze the bulk as well as the surface non-electrostatic, ion specific effects.

The differential capacitance of the electric double layer is a response function that characterizes how much charge is stored at the interface of the Coulomb fluid and depends strongly on the applied potential \cite{Bockris}. It has been proposed recently that the details of this dependence and especially the deviations from the straightforward Gouy-Chapman prediction, displaying a symmetric surface potential dependence with a minimum at the vanishing potential of zero charge (PZC), can be attributed at least in part to non-electrostatic ion-ion interactions and that a simple generalization of the PB theory to either the case of a symmetric or asymmetric Coulomb lattice gas can qualitatively account for the observed double hump dependence on the surface potential \cite{Kornyshev-rev, Leermalers2009, Woodward2010, Andelman2015-capacitance, Kornyshev2018}. While this ion-ion packing effect can be observed in the surface response function such as the differential capacitance \cite{TianyingYan1, TianyingYan2, Seddon2018}, this does not really shed any light on the effects of the genuine ion-surface non-electrostatic interactions, in particular on the specific ion adsorption \cite{Goodwin,Uematsu}. And if one acknowledges the non-electrostatic effects between the ions in the vicinity of the surface as well as between the ions and the surface itself, the question arises as to which would be predominant and how do they differ.

With all this in mind, a generalization of the mean-field approach proposed by Maggs and Podgornik \cite{maggs2016general} will be derived, that will take into account the ion-ion as well as ion-surface non-electrostatic effects on an equal footing, being based on the bulk and surface equations of state in the absence of electrostatic interactions. This approach will then be applied to the calculation of the differential capacitance for several models of the specific ion-surface non-electrostatic interactions. While again this local thermodynamics based approach cannot have quantitative ambitions and does not supersede the MC and MD results in any quantitative way, it can certainly provide a valuable conceptual and organizing tool, as has indeed always been the case with thermodynamics.

\section{Volume term and generalized PB theory}

Charging up an isothermal ($T = const$) binary mixture of particles, that is described by the free energy $f(c_1, c_2)$ in the uncharged state, or equivalently with the equation of state $p(\mu_1, \mu_2)$, where $p$ is the (osmotic) pressure, $c_1, c_2$ are concentrations and $\mu_1, \mu_2$ are the two chemical potentials,  leads, within the local thermodynamics approximation, to the inhomogeneous thermodynamic potential of the form \cite{maggs2016general} 
\begin{align}
  {\cal F}_V[c_1, c_2, {\bf D} ] = \int_V\!\!\!d^3{\bf r} \left( f(c_1, c_2) - \mu_1 c_1 - \mu_2 c_2\right)  + \nonumber\\
  + \int_V\!\!\!d^3{\bf r} \left( {\textstyle\frac12} \frac{{\bf D}^2}{\varepsilon} -
  \psi\left( \bnabla\cdot {\bf D} - e (Z_1 c_1 - Z_2
  c_2)\right)\right),
  \label{cdrtqwy}
\end{align}
where ${\bf D} = {\bf D}({\bf r})$ is the dielectric displacement field, $\varepsilon = \epsilon\epsilon_0$ with $\epsilon$ the relative dielectric permittivity, $Z_{1,2}$ are the valencies of the two charged species and $\psi = \psi({\bf r})$ is now the Lagrange multiplier field that ensures the local imposition of Gauss' law \cite{Maggs1}.  This expression can be written in an alternative form by invoking the thermodynamic identity $f(c_1, c_2) - \mu_1 c_1 - \mu_2 c_2 = - p(\mu_1, \mu_2),$ discarding the boundary terms that will be considered separately later, and minimizing with respect to $\bf D$, yielding the final form of the inhomogeneous thermodynamic potential
\begin{equation} {\cal F}_V [\psi] = - \!\!\!\int_V\!\!\!d^3{\bf r}
  \Big( {\textstyle\frac12} {\varepsilon} (\bnabla\psi)^2 + p(\mu_1 - e
    Z_1 \psi, \mu_2 + e Z_2 \psi)\Big).
      \label{gfiwq}
\end{equation}
This furthermore implies that for the fully charged system
\begin{equation}
 \frac{\partial f(c_1, c_2)}{\partial c_{1,2}} =  \mu_{1,2} \mp eZ_{1,2} \psi,
 \label{ext1}
\end{equation}
where the r.h.s. are nothing but the electrochemical potentials. 

The derivation of Eq.~(\ref{gfiwq}) proceeded entirely on the mean-field level. By assumption, the inhomogeneous case is described on the {\sl local thermodynamic} approximation level, so that the inhomogeneity features solely via the coordinate dependence of the densities, but the form of the
thermodynamic potential remains the same as in the bulk.

Denoting the free energy density corresponding to the volume free energy Eq. \ref{gfiwq} as $f_V(\psi, \bnabla \psi)$, the corresponding volume Euler-Lagrange equation is then given by
\begin{equation}
\bnabla \frac{\partial f_V (\psi, \bnabla \psi)}{\partial \bnabla \psi} - \frac{\partial f_V ((\psi, \bnabla \psi))}{\partial  \psi} = 0
\end{equation}
implying that
\begin{equation} 
{\varepsilon} \bnabla^2 \psi =  \frac{\partial p(\mu_1 - e Z_1 \psi, \mu_2 + e Z_2 \psi)}{\partial \psi}.
\label{ncbfgjkwk}
\end{equation}
This form leads to a whole slew of generalized Poisson-Boltzmann equations, depending on the model of the uncharged system as encoded by its equation of state, $p(\mu_1, \mu_2)$. Recently the generalized PB equations were specifically considered in the case of an ideal two-component gas, with concentrations $c_1$ and $c_2$, that leads to an equation of state 
\begin{equation}
p(c_1, c_2) = k_B T(c_1 + c_2),
\label{PBform}
\end{equation}
with Eq. \ref{ncbfgjkwk} being then identical to the standard PB equation. A generalization of the PB equation was then obtained for the binary, symmetric lattice-gas as well as an asymmetric variant derived from the Flory-Huggins lattice level approximation and the Carnahan-Starling asymmetric binary mixture.  In all these generalizations each uncharged equation of state thus produces an associated generalized PB equation  \cite{maggs2016general}. 

\section{Surface term and generalized charge regulation theory}

A similar approach will now be proposed also in the case of a surface phase in equilibrium with the bulk. Standardly for many processes taking place at the interface the introduction of surface excess quantities, that relies on an infinitely thin surface transition layer, is a reasonable approximation in thermodynamics \cite{Widom} as well as in the ionic solution theory \cite{Markovichsurfacetension}, though in simulations it is of course important to keep a finite thickness of the surface layer described by different material properties then the bulk as in the recent analysis of the zeta potential \cite{Uematsu}.

In the case of a bounding surface one can first decompose the total number of particles into the bulk and the surface part
\begin{equation}
N_{1,2} = \int_V d^3{\bf r} ~c_{1,2}  + \oint_S d^2{\bf r} ~c_{S1,S2},
\end{equation}
so that the surface excess concentration for the two components is $c_{S1,S2}$. The integral $\oint_S d^2{\bf r}$ runs over all the boundary surfaces of the system.

One assumes that the surface can interact with both types of charge species, i.e., $1,2$, an assumption that can be easily generalized or restricted. This leads to the following contribution of the total surface free energy 
\begin{eqnarray}
 &&{\cal F}_S[c_{S1}, c_{S2}, {D_n}] = \oint_S\!d^2{\bf r} \Big( f_S(c_{S1}, c_{S2})\!\! -\!\! \mu_1 c_{S1}\!\! -\!\! \mu_2 c_{S2}\Big)  - \nonumber\\
 && -  \oint_S d^2{\bf r}~ \psi_S \Big( D_n + (e c_{S0} - eZ_1 c_{S1} + eZ_2 c_{S2}) \Big),
  \label{cdrtqwyS}
\end{eqnarray}
where by assumption that $"1"$ is positively charged and "2" and "0" negatively, while $\psi_S$ is the surface Lagrange multiplier field imposing the Gauss' law in the form of the normal component of the dielectric displacement vector  ${D_n}$. The local normal is defied so as to point into the region that contains the charged mobile particles.  The term $e c_{S0}$ pertains to the possible fixed charge on the surface, independent of the local concentrations of adsorbing species $1,2$.

The complete thermodynamic potential is then composed of the bulk contribution Eq. \ref{cdrtqwy} and the surface contribution Eq. \ref{cdrtqwyS} as
\begin{eqnarray}
  & & {\cal F}[c_1, c_2, {\bf D}, D_n]  = {\cal F}_V[c_1, c_2, {\bf D}] + {\cal F}_S[c_{S1}, c_{S2}, {D_n}]. 
 \label{AnsatzS}
\end{eqnarray}
Minimization of this thermodynamic potential w.r.t. $\bf D$ leads to the following Euler-Lagrange equations.
\begin{equation}
{\bf D} = - \varepsilon \bnabla \psi \qquad {\rm and} \qquad \psi_{\partial V} = \psi_S.
\end{equation}
The last equation signifies that the surface Lagrange field $\psi_S$ is equal to the volume Lagrange field $\psi$ evaluated at the surface \cite{Schwinger}. This follows after realizing that the direction of the normal in the Gauss theorem and the direction of the normal defined above are opposite. Inserting these identities back into Eq. \ref{AnsatzS} one remains with
\begin{eqnarray}
  & & {\cal F}[\mu_1, \mu_2, \psi]  = \nonumber\\
  &=& \int_V\!\!\!d^3{\bf r} \left( - \frac{1}{2}\varepsilon(\bnabla \psi)^2 - p(\mu_1 - e Z_1 \psi, \mu_2 + e Z_2 \psi) \right)  + \nonumber\\
  &+& \oint_S\!\!\!d^2{\bf r}~ \Big( - \psi_S e c_{S0} + \Sigma(\mu_1 - e Z_1 \psi_S, \mu_2 + e Z_2 \psi_S)\Big).\nonumber\\
  ~
\label{nfgjkweh}
\end{eqnarray}
The thermodynamic equilibrium state then follows directly from the thermodynamic potential now dependent only on the local electrostatic potential. Above, the surface tension ({\sl surface equation of state}) is introduced  as \cite{Widom}
\begin{eqnarray}
f_S(c_{S1}, c_{S2}) - &&(\mu_1 - e Z_1 \psi_S) c_{S1} - (\mu_2 + e Z_2 \psi_S) c_{S2} = \nonumber\\ &&\Sigma(\mu_1 - e Z_1 \psi_S, \mu_2 + e Z_2 \psi_S),
\label{surf}
\end{eqnarray}
implying furthermore that the surface equilibrium of the fully charged system is given by 
\begin{equation}
 \frac{\partial f_S(c_{S1}, c_{S2})}{\partial c_{S1,S2}} =  \mu_{1,2} \mp eZ_{1,2} \psi_S,
  \label{ext3}
\end{equation}
where the r.h.s. are now the surface electrochemcial potentials.

Introducing first the full surface free energy density corresponding to Eq. \ref{cdrtqwyS} as $f_S(\psi_S)$, the surface part of the Euler-Lagrange equations becomes
\begin{equation}
\frac{\partial f_V (\psi, \bnabla \psi)}{\partial \bnabla \psi} \cdot {\bf n}  + \frac{\partial f_S(\psi_S)}{\partial \psi_S} = 0.
\end{equation}
Here one needs to be careful about the proper definition of the sign of the normal to the bounding surface.  Notably the l.h.s. term of this equation stems from the volume part of the electrostatic energy in Eq. \ref{gfiwq}, while the r.h.s. is derived purely from the surface part. The final form of this equation then assumes the form
\begin{equation} 
{\varepsilon} \left(\bnabla \psi_S \cdot {\bf  n}\right) =  -e c_{S0} + \frac{\partial \Sigma\left( \mu_1 - e Z_1 \psi_S, \mu_2 + e Z_2 \psi_S\right)}{\partial \psi_S} .
\end{equation}
This is then just the surface equivalent of the Poisson-Boltzmann equation for the volume part of the free energy, and actually constitutes its boundary condition that notably depends on the value of the boundary potential, $\psi_S$, and thus the above equation constitues a generalized {\sl charge regulation boundary condition}, where the surface charge and surface potential have to be determined self-consistently. Each model of the surface tension $\Sigma(\mu_1, \mu_2)$ leads to a different surface charge regulation condition.

\section{Bulk and surface Euler-Lagrange equations}

The volume and surface Euler-Lagrange equations can now be redressed in a form that makes them recognizable for what they are, {\sl i.e.}, the {\sl Poisson equation} and the {\sl Gauss equation}. Invoking first  the {\sl Gibbs-Duhem relation} $$c_{1,2} = \frac{\partial p}{\partial \mu_{1,2}}$$ one can derive the volume part of the Euler-Lagrange equations as the Poisson equation
\begin{align}
  \frac{\partial p(\mu_1 - e Z_1 \psi, \mu_2 + e Z_2
  \psi)}{\partial \psi} =& - e Z_1  \frac{\partial p}{\partial \mu_{1}} + e Z_2 \frac{\partial p}{\partial \mu_{2}} =\nonumber\\
                         & = - e (Z_1c_1 - Z_2c_2) = -\rho. 
                           \label{befjkw}
\end{align}
where $\rho$ is the local charge density.  Furthermore, invoking the fact that $c_{S1, S2}$ are the surface excess concentrations, the {\sl Gibbs adsorption isotherm} implies that
\begin{equation}
c_{S1} = - \frac{\partial \Sigma}{\partial \mu_{1}} \quad {\rm and} \quad c_{S2} = - \frac{\partial \Sigma}{\partial \mu_{2}}
\label{gibbsiso}
\end{equation}
and allows one to rewrite the surface part of the Euler-Lagrange equations as
\begin{eqnarray}
\frac{\partial \Sigma\left(\mu_1 - e Z_1 \psi_S, \mu_2 + e Z_2 \psi_S\right)}{\partial \psi_S} && = - eZ_1 \frac{\partial \Sigma}{\partial \mu_{1}} + eZ_2 \frac{\partial \Sigma}{\partial \mu_{2}} = \nonumber\\
& & + eZ_1 c_{S1} - eZ_2 c_{S2}.
\end{eqnarray}
The final form of the full set of the Euler-Lagrange equations is then
\begin{eqnarray} 
&&{\varepsilon} \bnabla^2 \psi =  \frac{\partial p(\mu_1 - e Z_1 \psi, \mu_2 + e Z_2 \psi)}{\partial \psi} = \nonumber\\
&& - e (Z_1c_1 - Z_2c_2) = - \rho(\psi),
\label{bulkPB}
\end{eqnarray}
and
\begin{eqnarray} 
&&{\varepsilon} \left(\bnabla \psi_S \cdot {\bf  n}\right) = \frac{\partial \Sigma\left(\mu_1 - e Z_1 \psi_S, \mu_2 + e Z_2 \psi_S\right)}{\partial \psi_S} = \nonumber\\ 
&& - e c_{S0} + eZ_1 c_{S1} - eZ_2 c_{S2}= \sigma_C(\psi_S).
\label{surfaceBC}
\end{eqnarray}
where $\sigma_C$ is the total surface charge density.   Also, one observes that using a surface lattice gas expression for $\Sigma(\mu_1)$ would lead straight to the Ninham-Parsegian charge regulation boundary condition \cite{ninham} and other more complicated expressions if the lattice gas {\sl Ansatz} is changed \cite{0295-5075-113-2-26004}.

Another way to proceed from the complete thermodynamic potential of the system Eq. \ref{nfgjkweh} is to rewrite it first as
\begin{eqnarray}
  & & {\cal F}[c_1, c_2, \psi; c_{S1}, c_{S2}]  = \nonumber\\
  && - \int_V\!\!\!d^3{\bf r} ~\Big( \frac{1}{2\varepsilon} (\bnabla \psi)^2 + p\left(\mu_1 - e Z_1 \psi, \mu_2 + e Z_2 \psi\right) \Big) + \nonumber\\
&+& \oint_S\!d^2{\bf r}~ \left( e Z_1  c_{S1} - e Z_2  c_{S2} -  e c_{S0} \right)\psi_S + \nonumber\\
&+&  \oint_S\!d^2{\bf r}~\Big( f_S(c_{S1}, c_{S2})  -\mu_1c_{S1} -  \mu_2 c_{S2}\Big).
\label{nkcajwgscfkj1}
\end{eqnarray}
The first two lines now represent the thermodynamic potential of a system with surfaces charge density $\sigma_S(c_{S1}, c_{S2}) $ given by 
\begin{eqnarray}
\sigma_S(c_{S1}, c_{S2}) &=&  \left( e Z_1  c_{S1} - e Z_2  c_{S2} -  e c_{S0} \right) = \nonumber\\
&=& \sigma(c_{S1}, c_{S2}) -  e c_{S0}.\label{vfehjw}
\end{eqnarray} 
$\sigma$ defined above will be used later.

An explicit form for this electrostatic part of the thermodynamic potential can then be written with the Casimir charging formula \cite{Verwey} 
which allows one to write the full thermodynamic potential of the system in the form of a surface integral 
\begin{eqnarray}
\!\!\!& & {\cal F}[\sigma_S(c_{S1}, c_{S2}), c_{S1}, c_{S2}]  = \nonumber\\
  & &  \oint_S d^2{\bf r}~\Big(\int_0^{\sigma_S}\!\!\!\psi_S(\sigma) d\sigma + 
f_S(c_{S1}, c_{S2})  -\mu_1c_{S1} -  \mu_2 c_{S2} \Big), \nonumber\\ 
\label{nkcajwgscfkj2}
\end{eqnarray}
where the surface charge density $\sigma_S$ is given by Eq. \ref{vfehjw}. The surface Euler-Lagrange equations can then be obtained directly from Eq. \ref{nkcajwgscfkj2} as a variation with respect to the surface concentrations of the ions, {\sl viz.}, 
\begin{eqnarray}
\frac{\partial f_S(c_{S1}, c_{S2})}{\partial c_{S1}} &=& \mu_1 - eZ_1 \psi_S =  \mu_{S1} \nonumber\\
\frac{\partial f_S(c_{S1}, c_{S2})}{\partial c_{S2}} &=& \mu_2 + eZ_2 \psi_S = \mu_{S2},
\label{Marcus}
\end{eqnarray}
which of course coincide with the previously derived Eq. \ref{ext3}. Here $ \mu_{S1,S2}$ have been furthermore introduced explicitly as surface electrochemical potentials. These two equations represent the inverted Ninham-Parsegian boundary condition \cite{ninham} since the latter states the dependence of the surface charge density on the surface potential, while the above two equations describe the dependence of the surface potential on the surface (charge) density. 

In the form derived above, Eqs. \ref{Marcus} actually reduce to the formula derived by Marcus for the dissociation equilibrium of a stiff polyelectrolyte chain \cite{Marcus} in the context of titration phenomena. It could just as well be seen as a derivation of the charge regulation theory in the same context.

\section{Surface association/dissociation Models }

An analysis along the same lines as for the bulk term in Ref. \cite{maggs2016general} can now be pursued also for the surface free energy, $f_S(c_{S1},c_{S2})$ in Eq. \ref{surf}, or equivalently the surface tension $ \Sigma(\mu_1, \mu_2)$. In general, as demonstrated below, all non-trivial forms of the surface free energy lead to charge regulation models related to different models of surface-ion solution interactions \cite{borkovec2001ionization}. 

Several illuminating models will now be introduced just to illustrate the above general analysis.

\subsection{No specific surface-ion interactions}

In this trivial case $f_S(c_{S1},c_{S2}) = 0$, and the boundary condition Eq. \ref{surfaceBC} assumes its standard form of a fixed surface charge
\begin{equation} 
{\varepsilon} \left(\bnabla \psi_S \cdot {\bf  n}\right) =  - e c_{S0},
\end{equation}
most commonly used in the simplified version of the DLVO theory \cite{Verwey}.

\subsection{Single site dissociation model}

The surface dissociation processes is assumed to be of the type $\rm A^{-} + B^{+} \rightleftharpoons AB,$ where $\rm AB$ is the neutral complex between the solution ion $\rm A^{-}$ and the surface fixed charge moiety $\rm B^{+}$, with the equilibrium constant $K$ for the dissociation process. By assumption, in this case the ion that is exchanged between the surface and the bulk does not compete with any other ion.
 
\subsubsection{van't Hoff adsorption isotherm}

Assume first that one of the mobile ion types, say "1", adsorbs onto the surface and while adsorbed, acts as an ideal particle confined to the surface. The adsorbed phase is thus a dilute phase of mobile particles with full translational degrees of freedom along the adsorbing surface. In this case
\begin{eqnarray}
\label{Langmuir0}
&&f_S(c_{S1}) = - \alpha c_{S1} + k_B T c_{S1}\big( \ln (c_{S1} a^2) - 1\big),
\end{eqnarray} 
where $\alpha$ is the chemical potential  change for the adsorption process and $a$ is the typical size of the surface site. Writing Eq. \ref{ext3} in the form appropriate for this case one obtains
\begin{equation}
 \frac{\partial f_S(c_{S1})}{\partial c_{S1}} =  \mu_{1} - eZ_{1} \psi_S = - \alpha + k_BT \ln{(c_{S1} a^2)}
\end{equation}
so that
\begin{equation}
c_{S1} = \frac{1}{a^2}e^{\beta (\mu_{1} + \alpha - eZ_{1} \psi_S)} =  \frac{K}{a^2}e^{\beta (\mu_{1} - eZ_{1} \psi_S)}.
\end{equation}
Inserting this expression back into the Legendre transform of the surface free energy one obviously remains with
\begin{equation}
\Sigma\left(\mu_{1} - eZ_{1} \psi_S \right) = - k_B T c_{S1} = - \frac{k_B T}{a^2}  e^{\beta (\mu_{1}  +  \alpha - eZ_{1} \psi_S)}.
\end{equation}
One should note here that the r.h.s. of the above equation can be obtained from the grand canonical partition function of an {\sl ideal surface gas} in the form
\begin{equation}
\Sigma(\mu ) = - \frac{k_BT}{a^2} \log{\Xi_S} \qquad {\rm with} \qquad {\Xi_S} = e^{(e^{\beta \alpha} ~e^{\beta \mu})} 
\end{equation}
as a function of the surface electrochemical potential $\mu \equiv \mu_{1}  - eZ_{1} \psi_S$, where the $ \alpha$ parameter can be regarded as the chemical potential change for the association/dissociation (adsorption/desorption) process, so that the equilibrium constant  for this process is $K = e^{\beta \alpha} $. The surface excess density then follows from the Gibbs adsorption isotherm Eq. \ref{gibbsiso}. This model was used in \cite{MarkovichJCP} to analyze the effects of ionic specificity on the surface tension of electrolyte interfaces.

The boundary condition in this case follows from Eq. \ref{surfaceBC} as
\begin{eqnarray} 
&&{\varepsilon} \left(\bnabla \psi_S \cdot {\bf  n}\right) = - e c_{S0} + \frac{\partial \Sigma\left(\mu_1 - e Z_1 \psi_S\right)}{\partial \psi_S} = \nonumber\\ 
&& - e c_{S0} + \frac{eZ_1}{a^2} e^{\beta (\mu_{1} + \alpha - eZ_{1} \psi_S)} = - e c_{S0} +\sigma(\psi_S).
\end{eqnarray}

The van't Hoff adsorption isotherm is relevant when the adsorbed phase is mobile and the adsorbed molecules have full translational freedom along the surface, but are confined in the direction perpendicular to the surface \cite{Hill}. In this case the number of adsorbed particles is always greater then in the case when they are immobile on the surface, as is the case in the next adsorption model.

\subsubsection{Langmuir adsorption isotherm}

Now assume that again only ions of one type, "1", can interact with the positively charged {\em fixed surface sites}. As these sites become occupied there is less and less available positions for the particle adsorption just like in the standard lattice gas case \cite{Hill}.  One can therefore start with the free energy of the surface lattice gas in the form 
\begin{eqnarray}
\label{Langmuir1}
&&f_S(c_{S1}) = - \alpha c_{S1} + \nonumber\\
&& + k_B T \Big( c_{S1}\ln{(c_{S1} a^2)} + \left(1-{(c_{S1} a^2)} \right)\ln\left(1-{(c_{S1} a^2)} \right)\Big), \nonumber\\
~
\end{eqnarray}
where $\alpha$ is again the chemical potential  change for the dissociation process.  In this case, a derivation in complete analogy with the previous one, yields
\begin{eqnarray}
\Sigma\left(\mu_{1} - eZ_{1} \psi_S \right)&=& \frac{k_B T}{a^2} \ln{(1 - c_{S1}a^2)} = \nonumber\\
& &\!\!\!- \frac{k_B T}{a^2}  \ln{\left( 1 + e^{\beta (\mu_1 +  \alpha - eZ_{1} \psi_S)}\right)},\nonumber\\
~
\label{eq31}
\end{eqnarray}
with
\begin{equation}
c_{S1}a^2 = \frac{e^{\beta (\mu_{1} + \alpha - eZ_{1} \psi_S)}}{1 + e^{\beta (\mu_{1} + \alpha - eZ_{1} \psi_S)}} = \frac{K e^{\beta (\mu_{1}  - eZ_{1} \psi_S)}}{1 + K e^{\beta (\mu_{1} - eZ_{1} \psi_S)}}.
\end{equation}
The form Eq. \ref{eq31} has been derived by a different route already by Chan and Mitchell \cite{chan1983}, as well as later by Biesheuvel et al. \cite{biesheuvel2004electrostatic, BiesheuvelVeen}.  Again, one notes that the r.h.s. of Eq. \ref{eq31} is nothing but the grand canonical partition function of a surface lattice-gas  in the form
\begin{equation}
\Sigma\left(\mu\right) =  - \frac{k_BT}{a^2} \log{\Xi_S} \qquad {\rm with} \qquad {\Xi_S} = {\left(1 + e^{\beta \alpha} {e^{\beta \mu}}\right)}.
\label{onelang-1}
\end{equation}
The surface excess concentration is then obtained from $\Sigma\left(\mu_{1} - eZ_{1} \psi_S\right)$, i.e., the surface tension as a function of surface electrochemical potential by applying the Gibbs isotherm Eq. \ref{gibbsiso}, yielding the boundary condition  as 
\begin{eqnarray} 
&&{\varepsilon} \left(\bnabla \psi_S \cdot {\bf  n}\right) = - e c_{S0} + \frac{\partial \Sigma\left(\mu_1 - e Z_1 \psi_S\right)}{\partial \psi_S} = \nonumber\\ 
&& - e c_{S0} + \frac{eZ_1}{a^2} \frac{e^{\beta (\mu_{1} + \alpha - eZ_{1} \psi_S)}}{1 + e^{\beta (\mu_{1} + \alpha - eZ_{1} \psi_S)}}, 
\end{eqnarray}
or alternatively
\begin{eqnarray} 
&&{\varepsilon} \left(\bnabla \psi_S \cdot {\bf  n}\right) = - e c_{S0} + {eZ_1} c_{S1}(\psi_S).
\end{eqnarray}
If one furthermore assumes that $Z_1/a^2 = c_{S0}$, as will be in what follows, then one remains with
\begin{equation}
{\varepsilon} \left(\bnabla \psi_S \cdot {\bf  n}\right) = - {\textstyle\frac12}e c_{S0} + {\textstyle\frac12}e c_{S0}\tanh{{\textstyle\frac12}\beta (\mu_{1} + \alpha - eZ_{1} \psi_S)}.
\end{equation}
It is straightforward to see that this form is completely equivalent to the standard charge regulation boundary condition of Ninham and Parsegian \cite{ninham}.

\subsection{Many site dissociation model}

The surface dissociation equilibrium is now assumed to be of the type
$B_nA  \rightleftharpoons  A^- + n B^+$, with the equilibrium constant $K_n$ for each of the dissociation processes. Here, different species $B_nA$ will be considered, dependent on the number of dissociable ions but independent of the locations of the bound ions, only their number being relevant. On the connection with a molecular point of view see the discussion in Borkovec, Jonsson and Koper \cite{borkovec2001ionization}.

Generalizing the Langmuir adsorption isotherm to the case of many equivalent, independent, and distinguishable site dissociation processes, one can introduce the partition function $q(s)$ as the site partition function when $s$ molecules are bound to the site, where $0 < s < N$. The grand canonical partition function for such a system has been derived in the form  \cite{Hill, borkovec2001ionization}
\begin{eqnarray}
\Xi_S &=& \left( q(0) + q(1) {e^{\beta \mu}} + \dots +  q(N) {e^{N\beta \mu}}\right) = \nonumber\\
&& \sum_{m=0}^{N} q(m) e^{m\beta \mu}.
\label{onelang-2}
\end{eqnarray}
Clearly for a two site dissociation process $q(0) = 1$ and $q(1) = e^{\beta \alpha}$, giving back Eq. \ref{onelang-1}. The connection with the {\sl binding polynomial} is given by $q(n)/q(0) = K_n$, where $K_n$ are the standard equilibrium constants \cite{Hill}.

\subsubsection{Adsorption to pairs of sites}

An interesting example of a many site dissociation process is a system of {\sl independent pairs} of sites. Here $q(1)$ is the partition function for a molecule bound to the site of type 1, and $q(2)$ to a site of type 2. When both sites of a pair are occupied one furthermore assumes that the two bound particles interact with an non-electrostatic interaction energy $u$. Otherwise, there are no (non-electrostatic) interactions between the adsorbed particles and the sites are now pairs with  $s = 0, 1, 2$. The partition function Eq. \ref{onelang-2} then turns out as
\begin{equation}
\Xi_S(\mu) = \left( 1 + (K(1)+K(2)) {e^{\beta \mu_1}} + K(1)K(2) {e^{-\beta u}} {e^{2 \beta \mu_1}}\right).
\label{onelang-3}
\end{equation}
Clearly if there is no (non-electrostatic) interaction energy between the adsorbed ions, {\sl i.e.}, $u = 0$, then the adsorption processes are uncoupled and the partition function is reduced to a product. In the opposite limit of strong repulsive interaction the partition function becomes, {\sl i.e.}, equivalent to a single site adsorption. 

Introducing now the surface adsorption isotherm
\begin{equation}
\Sigma\left(\mu_1\right) =  - \frac{k_BT}{a^2} \log{\Xi_S},
\end{equation}
the boundary condition then follows as
\begin{eqnarray} 
{\varepsilon} \left(\bnabla \psi_S \cdot {\bf  n}\right) &=& - e c_{S0} + \frac{\partial \Sigma\left(\mu_1 - e Z_1 \psi_S\right)}{\partial \psi_S} = \nonumber\\
&& - e c_{S0} + \sigma(\psi_S),
\end{eqnarray}
or equivalently as the charge regulating boundary condition
\begin{eqnarray} 
&&{\varepsilon} \left(\bnabla \psi_S \cdot {\bf  n}\right) = - e c_{S0} + \nonumber\\
&& + \frac{e Z_1}{a^2} \frac{(K(1)+K(2)) + K(1)K(2) {e^{-\beta u}} e^{2\beta \mu_{S1}} }{ \left( 1 + (K(1)+K(2)) {e^{\beta \mu_{S1}}} + K(1)K(2) {e^{-\beta u}} {e^{2 \beta \mu_{S1}}}\right)} \nonumber\\
~
\end{eqnarray}
with the surface electrochemical potential $\mu_{S1}$ defined as before. Assuming again for simplicity $Z_1/a^2 = c_{S0}$, and depending on the value of the non-electrostatic pair interaction energy $u$, the above boundary condition leads to two limiting forms: for $\beta u \gg 0$ and $\beta u \ll 0$. 

The charge regulation conditions in these two limits can be understood as indicating that for large repulsion between the adsorbing particles the point where the charge changes most, or has the highest surface capacitance for a give chemical potential $\mu_1$, is given by $eZ_{1} \psi_S = \ln{(K(1)+K(2))}$, for the first limit, while it equals $eZ_{1} \psi_S = -{\textstyle\frac12} \beta u +  {\textstyle\frac12}\ln{(K(1)K(2))}$ for the second one.


\section{Surface capacitance.}

\subsection{General theory}
The definition of surface capacitance $\cal C$ proceeds from the total thermodynamic potential of the  system with set surface charge density as \cite{Bockris}
\begin{equation}
\frac{\delta^2 {\cal F}[\sigma_S]}{\delta \sigma_S^2}  =  {\cal C}^{-1}(\psi_S)
\end{equation}
where $\sigma_S$ is given by Eq. \ref{vfehjw}. Since the total thermodynamic potential  actually depends on ${\cal F}[\sigma_S(c_{S1}, c_{S2}), c_{S1}, c_{S2}]$, given by Eq. \ref{nkcajwgscfkj2}, this implies an inverse differential capacitance in the form
\begin{equation}
{\cal C}^{-1} = \frac{\partial \psi_S}{\partial \sigma_S} + \left( \frac{1}{(eZ_1)}\frac{\partial }{\partial c_{S1}} - \frac{1}{(eZ_2)} \frac{\partial }{\partial c_{S2}}\right)^2  f_S(c_{S1}, c_{S2}),
\label{finforfun}
\end{equation}
with $c_{S1, S2}$ a function of the surface potential via Eqs. \ref{Marcus}. The first term obviously represents the double layer capacitance, and the second term is the charge-weighted surface concentration susceptibility. From the above equation it follows straightforwardly that the double layer capacitance and the charge-weighted surface concentration susceptibilities act as serial capacitors. The total capacitance is thus set by the smaller of the two.

Since by the general properties of the Legendre transform \cite{Zia} 
\begin{equation}
  \sum_m \frac{\partial^2 \Sigma(\mu_1, \mu_2)}{\partial \mu_{i} \partial \mu_{m}} \frac{\partial^2 f_S(c_{S1}, c_{S2})}{\partial c_{Sm} \partial c_{Sk}} = \delta_{ik},
\end{equation}
where all the matrices are $2 \times 2$, it then follows also that
\begin{eqnarray}
{\cal C}^{-1}(\sigma_S) &=& \frac{\partial \psi_S}{\partial \sigma_S} + \nonumber\\
&& {\cal D}^{-1}\left( \frac{1}{(eZ_1)}\frac{\partial }{\partial \mu_{1}} + \frac{1}{(eZ_2)} \frac{\partial }{\partial \mu_2}\right)^2  \Sigma(\mu_{1}, \mu_{2}), \nonumber\\
~
\label{ncsbrkunb}
\end{eqnarray}
where ${\cal D} =  {\rm det}\frac{\partial^2 \Sigma(\mu_1, \mu_2)}{\partial \mu_{i} \partial \mu_{m}}$. In the second term the surface electrostatic potential enters via the surface electrochemical potential $\Sigma(\mu_1, \mu_2)  \longrightarrow \Sigma(\mu_1 - eZ_1\psi_S, \mu_2 + eZ_2 \psi_S)$. All the terms in the above expression are then functions of the surface electrostatic potential. The above relation depends of course on the assumption of the volume and surface term decomposition in the thermodynamic potential, Eq. \ref{AnsatzS}. 

\begin{figure}[t!]
\begin{center}  
\includegraphics[width=0.35\textwidth]{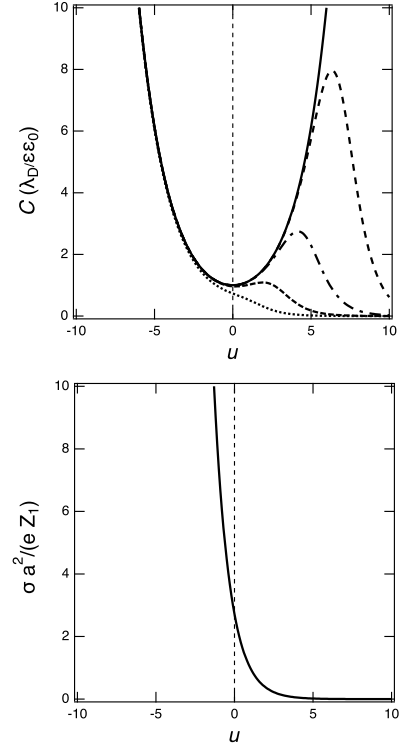} 
\caption{Dimensionless total surface capacitance $\tilde{\cal C} = {\cal C} (\lambda_D/\varepsilon\varepsilon_0)$ and dimensionless surface chareg density $\tilde\sigma_e$ for the van't Hoff model as a function of dimensionless surface potential $u =  eZ_1 \psi_S$ for $\mu_1 + \alpha = 1$ and $\chi = 0.0002, 0.005, 0.1, 1$ (from upper to lower curves). The double-layer capacitance ($\chi = 0$) is symmetric and shows no saturation effects. Only the positive dimensionless surface electrostatic potential shows surface adsorption effects.
\label{graph3}}
\end{center}
\end{figure}

The first term in ${\cal C}^{-1}(\sigma_S)$ is the standard inverse double-layer PB surface capacitance. It relies on the van't Hoff bulk equation of state, Eq. \ref{PBform}, for the two solution ions, based equivalently on the mean-field ideal gas entropy of the ions. One can however, derive related expressions also in the case of generalized PB equations, with more complicated equations of state. In this case, based on the completely general equation of state $p\left(\mu_1, \mu_2\right) $, one derives the modified Grahame equation \cite{maggs2016general} as
\begin{eqnarray}
&& \sigma_S(\psi_S) = \varepsilon \psi'(0) = \nonumber\\
&& = \pm \sqrt{2 \varepsilon} \Big(p\left(\mu_1 - e Z_1 \psi_S, \mu_2 + e Z_2 \psi_S\right) - p_0\Big)^{1/2},
\label{gengra}
\end{eqnarray}
where the overall sign depends on the sign of the surface charge, while $p_0$ is the bulk ion osmotic pressure corresponding to the model used for the equation of state of the uncharged fluid. From here it follows that the double layer capacitance for a generalized PB model is given by
\begin{eqnarray}
\frac{\partial \sigma_S(\psi_S)}{\partial \psi_S} &=& \sqrt{2 \varepsilon} \frac{\partial}{\partial \psi_S} \sqrt{p\left(\mu_1\!-\!e Z_1\!\psi_S, \mu_2\!+\!e Z_2\!\psi_S\right)\!-\!p_0}= \nonumber\\
&& \mp \sqrt{\frac{\varepsilon}{2}} \frac{\rho(\psi_S)}{\sqrt{p\left(\mu_1 - e Z_1 \psi_S, \mu_2 + e Z_2 \psi_S\right) - p_0 }}. \nonumber\\
~
\label{tyurewi}
\end{eqnarray}
Specializing again to the ideal gas form of the equation of state, Eq. \ref{PBform}, Eq. \ref{tyurewi} yields back the standard PB result
\begin{equation}
\frac{\partial \sigma_S(\psi_S)}{\partial \psi_S} = 
\frac{\varepsilon}{\lambda_D}\cosh{\frac{e\psi_S}{2 k_BT}},
\label{mhiorpt}
\end{equation}
obviously symmetric with respect to the zero of the surface potential. In what follows the capacitance will be analyzed  in units of the PB capacitance \cite{Kornyshev-rev} ${\varepsilon\varepsilon_0}/{\lambda_D}$ where $\lambda_D$ is the Debye screening length, which is given as $\lambda_D^2 = 1/(4\pi~\ell_B I)$ for a two component electrolyte. Here $I$ is the ionic strength and $\ell_B = e^2 \beta/(4\pi~\varepsilon\varepsilon_0)$ is the Bjerrum length.

The final form of the differential capacitance for a charge regulated model $\cal C$ is then given by Eq. \ref{finforfun} or equivalently by Eq. \ref{ncsbrkunb}, where the double layer capacitance is obtained in general from Eq. \ref{gengra}.

\begin{figure}[t!]
\begin{center}  
\includegraphics[width=0.35\textwidth]{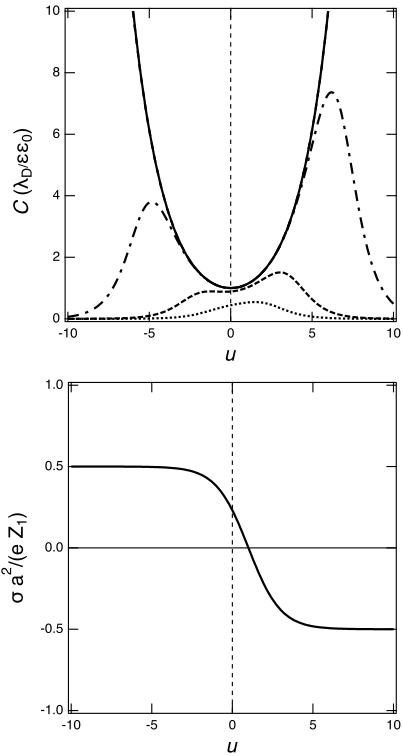} 
\caption{Dimensionless total surface capacitance $\tilde{\cal C} = {\cal C} (\lambda_D/\varepsilon\varepsilon_0)$ for the Langmuir model as a function of dimensionless surface potential $u =  eZ_1 \psi_S$ for $\mu_1 + \alpha = 1$ and $\chi = 0.000003, 0.001, 0.1, 1$ (from upper to lower curves). The double-layer capacitance ($\chi = 0$) is symmetric and shows no saturation effects. Both, the positive as well as the negative branch of the dimensionless surface electrostatic potential shows surface adsorption effects, which are asymmetric with respect to zero - a simple consequence to non-symmetric adsorption isotherm.
\label{graph4}}
\end{center}
\end{figure}

\subsection{Model results}

In case there is {\bf no specific ion-surface interaction} the total surface capacitance equals just the double layer capacitance and can be written as \cite{Kornyshev-rev}
\begin{equation}
\tilde{\cal C} = \cosh{\frac{e\psi_S}{2 k_BT}},
\label{mhiorpt1}
\end{equation}
where the dimensionless capacitance is $\tilde{\cal C} = {\cal C} \times (\lambda_D/{\varepsilon\varepsilon_0})$.

In the {\bf van't Hoff model} the following form of the inverse differential capacitance of the adsorption layer is obtained as
\begin{eqnarray}
 \frac{1}{(eZ_1)^2} \frac{\partial^2 f_S(c_{S1})}{\partial c_{S1}^2}&& =  \frac{1}{(eZ_1)^2} \left(\frac{\partial^2 \Sigma(\mu_{1})}{\partial \mu_{1}^2}\right)^{-1} = \nonumber\\
 && =\frac{k_BT~a^2}{(eZ_1)^2} ~e^{-\beta(\mu_1 + \alpha - eZ_1 \psi_S)}, 
\end{eqnarray}
and therefore
\begin{eqnarray}
\tilde{\cal C}^{-1} = \frac{1}{\cosh{\frac{e\psi_S}{2 k_BT}}} + \chi ~e^{-\beta(\mu_1 + \alpha - eZ_1 \psi_S)},
\end{eqnarray}
where $\chi = {\textstyle\frac12} \ell_{GC}/\lambda_D$ and $\ell_{GC} = 1/(2\pi \ell_B)~(e/|\sigma|)$ is the standard Gouy-Chapman length \cite{markovich2016charged}. The second term, stemming from the specific interaction with the surface, confers asymmetry to the total surface capacitance of the system. Numerical examples, see Fig. \ref{graph3}, illustrate the connection between the effective surface charge and the surface capacitance. Obviously the competition between the PB double layer capacitance and the contribution of the surface specific interactions leads to a highly non-monotonic capacitance, with the surface specific contribution dominating in the range of surface potentials where the surface charge $\sigma(\psi_S)$ changes most.

In the case of the {\bf Langmuir model} the adsorption isotherm is obtained similarly as
\begin{eqnarray}
&&\frac{1}{(eZ_1)^2}\frac{\partial^2 f_S(c_{S1})}{\partial c_{S1}^2} =  \frac{1}{(eZ_1)^2} \left(\frac{\partial^2 \Sigma(\mu_{1})}{\partial \mu_{1}^2}\right)^{-1} = \nonumber\\
&=&  4 \frac{k_BT~a^2}{(eZ_1)^2} \cosh^{2}{{\textstyle\frac12}\beta(\mu_1 + \alpha - eZ_1 \psi_S)}, 
\end{eqnarray}
and thus
\begin{eqnarray}
\tilde{\cal C}^{-1} = \frac{1}{\cosh{\frac{e\psi_S}{2 k_BT}}} + 4 \chi \cosh^{2}{{\textstyle\frac12}\beta(\mu_1\!+\!\alpha\!-\!eZ_1 \psi_S)},
\end{eqnarray}
where $\chi$ has been defined above. Both, for the van't Hoff as well as the Langmuir isotherms, obviously the surface concentration susceptibility is not symmetric with respect to the origin of $\psi_S$, as is the case for the PB differential double layer capacitance, Eq. \ref{mhiorpt}. Numerical examples, see Fig. \ref{graph4}, again support the conclusion that the non-monotonicity and the asymmetry of the total capacitance is due to a tradeoff between the double layer contribution, exhibiting a minimium at $\psi_S=0$, and the surface concentration susceptibility exhibiting a single maximum at the largest change in the effective surface charge with the surface potential. 

\begin{figure}[t!]
\centering
\includegraphics[width=0.5\textwidth]{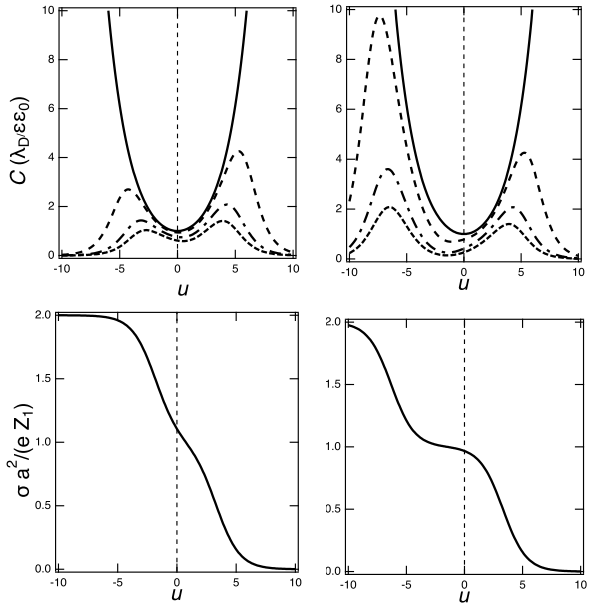} 
\caption{Dimensionless total surface capacitance $\tilde{\cal C} = {\cal C} (\lambda_D/\varepsilon\varepsilon_0)$ for the diprotic dissociation equilibrium model as a function of dimensionless surface potential $u =  eZ_1 \psi_S$ for $\mu_1 + \alpha = 1$. From right to left:  $\beta w = 0.5$, $pK_1-pH = 1$, $pK_2-pH= -1$ and $\chi = 0.01, 0.05, 0.1$ (from upper to lower curves); $\beta w = 5$, $pK_1-pH = 1$, $pK_2-pH= -1$ and $\chi = 0.01, 0.05, 0.1$ (from upper to lower curves). The equilibrium dissociation contstants are given as $K(1,2) = e^{\ln{10} ~(pK_{1,2}-pH)}$.
\label{graph5}}
\end{figure} 

For {\bf many site dissociation model} the surface concentration susceptibility can be derived as
\begin{eqnarray}
\frac{1}{(eZ_1)^2}\frac{\partial^2 f_S(c_{S1})}{\partial c_{S1}^2}  &=&  \frac{1}{(eZ_1)^2} \left(\frac{\partial^2 \Sigma(\mu_{1})}{\partial \mu_{1}^2}\right)^{-1} = \nonumber\\
&=& \frac{k_BT~a^2}{(eZ_1)^2} \Big(\left< m^2\right> - \left< m\right>^2\Big)^{-1} = \nonumber\\
&&=  \frac{(k_BT)^2~a^2}{(eZ_1)^2} \left(\frac{\partial \left< m\right>}{\partial \mu}\right)^{-1},
\end{eqnarray}
where
\begin{eqnarray}
\left< \dots \right> = \frac{\sum_{m=0}^{N} \dots q(m) e^{m\beta (\mu_1  - eZ_1 \psi_S)}}{\sum_{m=0}^{N} q(m) e^{m\beta (\mu_1  - eZ_1 \psi_S)}}
\end{eqnarray}
and is thus inversely proportional to the variance of the number of molecules bound to the surface sites. This leads to 
\begin{equation}
\tilde{\cal C}^{-1} = {\cosh^{-1}{\frac{e\psi_S}{2 k_BT}}} + \chi ~\Big(\left< m^2\right> - \left< m\right>^2\Big)^{-1},
\end{equation}
where again $\chi = {\textstyle\frac12} \ell_{GC}/\lambda_D$.

As an example of the many site dissociation model one can consider the diprotic dissociation equilibrium with $N=2$. In this case three phenomenological constants are needed, see Eq. \ref{onelang-3}, and the average number of surface bound molecules  is given by
\begin{eqnarray}
\left< m\right> &=&  \frac{(K(1)+K(2)) {e^{\beta \mu_1}} + 2 K(1)K(2) {e^{-\beta u}} {e^{2 \beta \mu_1}}}{1 + (K(1)+K(2)) {e^{\beta \mu_1}} + K(1)K(2) {e^{-\beta u}} {e^{2 \beta \mu_1}}},\nonumber\\
~
\end{eqnarray}
while 
\begin{eqnarray}
\left< m^2\right> &=& \frac{(K(1)+K(2)) {e^{\beta \mu_1}} + 4 K(1)K(2) {e^{-\beta u}} {e^{2 \beta \mu_1}}}{1 + (K(1)+K(2)) {e^{\beta \mu_1}} + K(1)K(2) {e^{-\beta u}} {e^{2 \beta \mu_1}}}.\nonumber\\
~
\end{eqnarray}
For the protonation reaction one can identify $e^{\beta \mu_1} K(1,2) = e^{\ln{10} ~(pK_{1,2}-pH)}$. 

Numerical examples for this case of surface dissociation process, see Fig. \ref{graph5}, show a more complicated picture due to the two processes of the surface adsorption/desorption characterized by two different equilibrium constants, as well as due to the interaction between the ions once adsorbed to the surface sites. The effective surface charge $\sigma(\psi_S)$ in this case exhibits two more or less pronounced - depending on the parameters of the model - regions of fast change with respect to $\psi_S$. Consequently the surface concentration susceptibility displays in general two unequal maxima at asymmetric positions with respect to zero of the surface potential that then modify the total surface capacitance now exhibiting two unequal, asymmetric local maxima.  

\section{Discussion and conclusions}

By following the general local mean-field thermodynamic approach appplied recently in the context of dense ionic liquids, where steric effects become relevant \cite{maggs2016general}, the mean-field theory of electrostatic double layers can be generalized to encompass bounding surfaces exhibiting active dissociation/association processes with solution ions, or in other words, exhibiting charge regulation. This can be done straightforwardly and as an outcome provides among other things also a description of the generalized surface capacitance of an electrified bounding surface that can be decomposed into the standard double-layer part and a modification stemming from the surface charge regulation.

In fact, the surface capacitance in the presence of active surface dissociation/association processes exhibits very different characteristics in comparison to its bare electrostatic double layer limit, valid for a surface that is completely passive with respect to the interactions with the solution ions. Any model of the specific solution ion-surface interactions, and several were analyzed without any attempt to be exhaustive, first of all reduces the total surface capacitance in the region of surface potentials where the active surface dissociation/association contribution to the capacitance is smaller then the bare electrostatic double layer limit, and in addition introduces noticeable assymmetry resulting from specific interactions of the different ion types with the surface, such as the steric saturation of the surface adsorption sites, thus eliminating the symmetric global minimum of the double layer capacitance at zero surface potential. 

In some respects these effects are similar to the {\sl bulk steric effects} of the ions in the vicinity of a passive charged surface explored in a series of recent works \cite{Andelman2015-capacitance,maggs2016general, Uematsu, Kornyshev-rev, TianyingYan1, TianyingYan2}, as they in general lead to an asymmetric double-hump dependence. There is however a fundamental difference between bulk steric effects and surface effects due to the ion boundary interactions. The latter differ from the steric effects in the bulk as they are associated with the density of dissociation/association sites, and not necessarily with the close packing of ions at the surface. The dissociation saturation and steric saturation effects can be well separated in concentration, allowing for a clear comparison of the two types of effects that depend either on the interaction with the surface or on the interaction between the ions. One is thus led to conclude that the packing considerations for surface charge regulation, that can set in at much smaller surface ion densities then those corresponding to bulk close packing, could be more relevant then the packing constraints for ions that are in the vicinity of the surface, but not actually adsorbed onto it, and are therefore bound to set in prior to the conditions where the bulk packing constraints would become relevant.
 
\section{Acknowledgments}

The author would like to acknowledge the support by the 1000-Talents Program of the Chinese Foreign Experts Bureau. He would also like to thank Prof. A.C. Maggs, Prof. D. Andelman, Prof. A.A. Kornyshev and Prof. Y. Levin for helpful comments.

\section{Bibliography}

\bibliography{rudip}

\end{document}